# Interplay of electro-thermo-solutal advection and internal electrohydrodynamics governed enhanced evaporation of droplets

**Vivek Jaiswal** and **Purbarun Dhar**

Department of Mechanical Engineering, Indian Institute of Technology Ropar,

Rupnagar–140001, India

*Corresponding author*:
E–mail: purbarun@iitrpr.ac.in; pdhar1990@gmail.com
Phone: +91–1881– 24–2173

## Abstract

The present article experimentally establishes the governing influence of electric fields on the evaporation kinetics of pendant droplets of conducting liquids. It has been shown that the evaporation kinetics of pendant droplets of saline solutions can be largely augmented by the application of an external alternating electric field. The evaporation behaviour is modulated by increase in the field strength as well as the field frequency. The classical diffusion driven evaporation model is found insufficient in predicting the improved evaporation rates. The change in surface tension due to field constraint is also observed to be insufficient for explaining the observed physics. Consequently, the internal hydrodynamics of the droplet is probed employing particle image velocimetry. It is revealed that the electric field induces enhanced internal advection, which is the cause behind the improved evaporation rates. An analytical model based on scaling approach has been proposed to understand the role of internal electrohydrodynamics, electro-thermal and the electro-solutal effects. Stability maps reveal that the electrohydrodynamic advection is caused nearly equally by the electro-solutal and electro-thermal effects and are the dominant transport mechanisms within the droplet. The model is able to illustrate the influence played by the governing thermal and solutal Marangoni number, the electro-Prandtl and electro-Schmidt number, and the associated Electrohydrodynamic number. The magnitude of the internal circulation can be well predicted by the proposed model, which thereby validates the proposed mechanism. The present revelations may find strong implications in droplet modulation using electric fields in micro and macroscale transport phenomena domains.

# 1. Introduction

Research and developmental activities on fundamentals and applied concepts of microfluidic systems has received significant attention in recent times due to the promise of such systems in the fields of biomedical engineering and healthcare, automotive industry, manufacturing sector, detection and quality control and several more. Among microscale fluidic domains, the droplets are an important paradigm which is responsible for the thermofluidic efficiency of several macroscale and microscale fluidic systems. Performance of applications such as agro-spraying, medicinal sprays and nebulizers, coating and spray-painting technologies, spray cooling systems, automotive engines and gas turbine systems, etc. are strongly dependent on the thermofluidic behaviour of sprays and their constituent droplets. Thereby, understanding of the hydrodynamics, thermal and phase interaction of droplets is essential for further improvement and efficient design of existing system and towards invoking ideas for development of new systems.

According to the prevalent regimes, droplets are generally classified into the sessile and the pendant droplets. A droplet rested in equilibrium on a substrate is termed as a sessile droplet, whereas a pendant droplet remains suspended in equilibrium (of gravitational and surface forces) from a small region of a surface (such as the tip of a needle). Evaporation dynamics of complex fluid droplets is a phenomenon which actively involves important hydrodynamic and thermal transport and species transport characteristics within the droplet. Several studies [1-4] have established two distinct modes of evaporation in case of sessile droplets viz. the constant contact angle (CCA) and constant contact radius (CCR) modes. Sessile droplet evaporation was studied in a pioneering work by Picknett and Bexon [1], where the dynamics of the contact line was given major focus. The role of the contact line dynamics was further discussed by Shanahan [5] in another thorough work. The roles played by relative humidity, thermal gradient across the substrate and the geometry of the droplet on the droplet evaporation kinetics have also been studied [6].

However, recently a shift of focus has occurred in the field of the droplet research, towards the dynamics of pendant droplets and related characteristics. This stems from the fact that pendent droplets are independent of surface interactions and hence the dynamics of constituent droplets of sprays and atomization processes can be modelled more accurately though the pendent approach. A pioneering work by Godsave [7] introduces the classical $D^2$ law for the evaporation of such pendent droplets. In more recent times, studies [8, 9] have shown that evaporation characteristic of such pendent droplets are modulated by tuning the interfacial properties employing salts or surfactants or particulate dispersions. Studies on the role of electromagnetic body forces on pendent droplet behaviour have also been reported (Rossow [10]). The enhanced evaporation dynamics of such droplets due to presence of ferro-convection within droplets of paramagnetic salts in presence of magnetic field has also been discussed in literature (Jaiswal et al. [11]). Electric body force has also been shown to influence dynamics and transport processes in droplets. Laroze [12] reports the role of body couple and Kelvin force that are encountered during evaporation of pendant droplet in an electric field environment. More recently, arrest and retardation of evaporation rate from pendent droplets specially orientated electric field has also been reported [13].

Electrohydrodynamics of droplets has received a fair share of attention owing to the rich physics involved and the applications, especially in microscale flow devices. Initial works in the field discussed the distortion and deformation of the droplet shape under the influence of electric field [14-19]. Theoretical predictions of the different modes of droplet deformation by combination of the classical Young-Laplace, Maxwell, Stokes and Laplace models have also been put forward [20]. The electric forces on droplets have led to the development of several microfluidic bio-medical devices, like therapeutic cell sorter, biological particle separator, drug testing rigs, gel electrophoresis and platelet segregators, to name a few. The electric field also imposes an additional strain field on the system, which leads to several electrohydrodynamic phenomena and applications, such as, electro-Leidenfrost effect, electro-wetting, electro-spinning and electro-propulsion. Studies by Digilov [21], Mugele and Baret [22] and Kang [23], Celestini and Kirstetter [24] and others focus on such electrohydrodynamics phenomena and applications. Gunji and Washizu [25] reported self-propelling droplets under the effect of electric field, and the electro-stick-slip phenomenon in sessile droplets was also introduced. The electro-stick-slip phenomenon represents a sliding action-reaction pair when a fluid droplet moves on a solid substrate due the interplay of an electric field and surface wettability.

The presence of electric field can also lead to the formation of a Taylor cone at microfluidic junctions can be used to generate well dispersed emulsions (Kim et al. [26]). Chabert et al. [27] reported the electro-coalescence of droplets due to alternating field effect. Hase et al. [28] explored the effects of a direct electric field on two identical droplets placed in between electrodes and reported a modulated droplet oscillation behaviour. A fascinating study reported the use of electrohydrodynamics actuated microfluidic jets towards cancer cell diagnostics using an alternating field (Jayasinghe et al. [29]). Evaporation or vaporization of droplets involves important applications. These include bio-medical applications (in spray based medicines, inhalers, nebulizers, blood culture, and DNA patterning, etc.), micro-manufacturing and processing methods like micro-lithography [30], ink-jet printing technology and in HVAC systems. The efficiency of fuel injection and ignition system of internal combustion engines [31-33] also depend on droplet evaporation kinetics. Droplet evaporation dynamics also play significant roles in the proper usage of pesticides, insecticides, fumigant [34, 35] and in microfluidic organ-on-chip devices [36].

A thorough survey of the literature shows that both evaporation dynamics of microscale droplets as well as droplet electrohydrodynamics has important implications and physics involved. Additionally, reports show that the presence of solvated ions enhances evaporation characteristics of pendant droplet [8], which furthers the motivation to understand the influence of an electric field on the evaporation behaviour of a conducting fluid pendent droplet. The present article experimentally and analytically explores the evaporation kinetics of a conducting fluid pendent droplet, and its associated electrohydrodynamics. Classical diffusion dominated evaporation models are found to be inadequate in predicting the enhanced evaporation rates of the droplet under electric field stimulus. Additionally, the change in surface tension and thermophysical properties was also found inadequate in explaining the enhanced liquid-vapour mass transport. Consequently, flow visualization employing Particle Image Velocimetry (PIV) was performed to quantify the internal electrohydrodynamics and its role in augmenting the evaporation has been discussed. A scaled analytical model to include the effects of electro-thermal and electro-solutal advection in the evaporation kinetics has been presented. The model predictions are observed to match well with the experimental findings. The present study may find strong implications in electrohydrodynamics modulated transport phenomena in droplets in microscale and macroscale flow regimes.

## 2. Materials and methodology

The present experiments were performed using a customized setup, which has been illustrated in figure 1. The setup consists of a precision droplet dispensing mechanism (Holmarc Opto-mechatronics, India) with a digitized controller. The dispenser can generate droplets accurate to 0.1 μL volume. The dispenser is mounted with a glass chromatography syringe (of capacity 100 μL), which has stainless steel needle (22 gauge) with a flat perforated end. For the present experiments, droplets of 15 ± 0.5 μL volume are suspended from the tip of the needle to form the pendants. The droplet volume is chosen to ensure that the droplet diameter is below the capillary length scale for water (~ 2.8 mm), thereby eliminating the role of gravity from the thermofluidic transport processes within the droplet. The evaporation process has been recorded at 10 fps using a charge-coupled device (CCD) sensor based camera, attached to a microscopic lens assembly (Holmarc Opto-mechatronics Pvt. Ltd.). A light emitting diode (LED) based illumination source (DpLED, China) was used as the backlight for the camera. The image acquisition was performed at a resolution of 1280 x 960 pixels. The complete setup was enclosed within an acrylic chamber to eradicate experimental condition fluctuations due to atmospheric conditions or human interference. A digital hygrometer and thermometer was used to measure humidity and temperature near the droplet (~5 cm away from the droplet, employing a local probe). For the whole set of experiments presented in this article, the temperature was recorded as 28 ± 3 $^{o}$C whereas humidity was recorded as 62 ± 3 %.

Fig.1 also illustrates the arrangement of the electrodes (two stainless steel plates, 20 mm x 10 mm). The electrodes are placed such that the electric field acts horizontally across the pendent droplet (inset fig.1). The droplet is positioned at the center of the electrodes to ensure field lines with minimal distortion. The gap between the electrodes is kept fixed at 4 mm. the electrodes are connected to a programmable AC power source (Aplab, India). It has a controllable field frequency range of 50 to 500 Hz with 0.1 Hz resolution, and controllable field strength ranging from 0 to 275 V, with 0.1 V resolution. Experiments with pure water droplet, at zero-field condition, but with the electrodes in position were performed to validate the customized setup. The control experiments were validated against the reports by Mandal and Bakshi [37] and good agreement with the literature was noted. Thereby, solely the presence of the electrodes was ascertained to be unimportant to the evaporation process.

The selection of salt solution has been made based on previous experiments by the present authors (Jaiswal et al. [8]). The report showed that sodium iodide (NaI) salt based droplets showed the best augmentation in the evaporation rates with respect to water, owing to the high solubility of the salt in aqueous systems. Solutions of anhydrous sodium iodide (Sigma Aldrich, India) of concentrations 0.05, 0.1 and 0.2 M were employed for the present study. The images of the evaporation process were processed in ImageJ, an open source image analysis tool. The droplet shape was fit to the Young-Laplace equation employing an open source macro and the instantaneous equivalent spherical diameter for the pendant droplet was deduced. Particle Image Velocimetry (PIV) has been done to qualitatively and quantitatively determine the internal hydrodynamics of the evaporating droplet. The salt solution has been seeded with neutrally buoyant fluorescent polystyrene particles (diameter ~10 μm, Cospheric LLC, USA). A continuous wave laser light source (Roithner Gmbh, Germany) is used and a plano-convex cylindrical lens has been used for generating a light sheet. The light sheet has been focused at the droplet mid-plane for the 2D velocimetry.

For a short time period of 3 minutes during the evaporation process, the LED backlight is dimmed off, and the laser sheet is used to perform the velocimetry. The wavelength of laser emission is 532 nm, and the power employed for the velocimetry is ~10 mW. The PIV is conducted during the initial stage of evaporation (within the first 5 minutes) to ensure that the evaporation induced increase in salt concentration does not lead to erroneous velocimetry. A CCD camera is used to capture the PIV frames at 30 fps, with a resolution of ~150 pixels/mm, and for duration of 90 seconds. An open source code, PIVlab, has been employed for the analysis. A cross-correlation algorithm, with four-pass window interrogation has been employed. Interrogation windows of 64, 32, 16 and 8 pixel widths have been used for the multi-pass analysis. Standard noise reduction algorithms have been used to improve the signal-to-noise ratio in the velocimetry analysis. Typical signal-to-noise ratios of 9-12 have been maintained for all processing. The velocimetry analysis is performed for 1000 consecutive frames, and the spatially averaged velocity for each frame is determined. The spatio-temporally averaged velocity field for all the frames is also determined, in order to obtain the mean velocity field within the droplet.

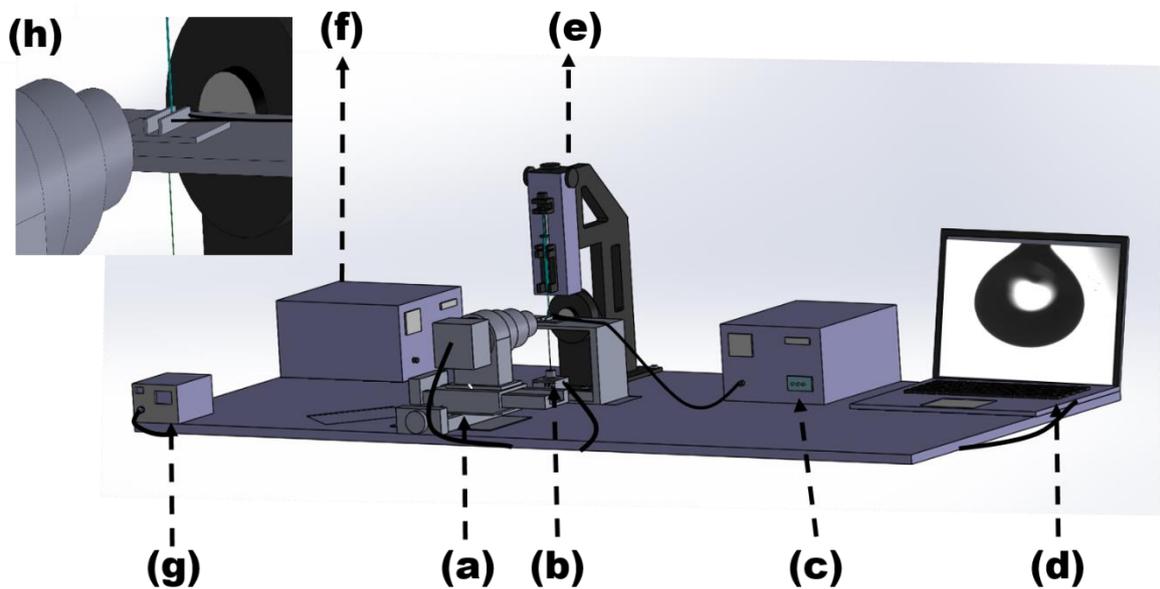

**FIG. 1:** Schematic of the experimental setup. It consists of (a) a CCD camera mounted on a two-axis movable frame and attached with a microscopic lens assembly(b) solid-state laser module mounted on the stand, with plano-convex cylindrical lens (not illustrated) for generating light sheet (c) programmable AC power source with frequency and field strength controller (d) data acquisition computer (e) precision droplet dispenser mechanism and backlight controlling module (f) droplet dispensing mechanism controller and programmer, and (g) power source and control unit for the laser. The inset (top-left corner) shows the zoomed in views of the arrangement of electrodes and the needle of the glass syringe. The whole experimental setup is enclosed within an acrylic chamber and mounted on a vibration-free table top during the study.

## 3. Results and Discussion

### 3. A. Evaporation characteristics under the influence of electric field

The evaporation behavior of water and salt solution droplets under zero-field has been used as the control case. The presence of electric field across the pure water droplet has been observed to have no influence over the evaporation dynamics (for the field strengths and field frequencies employed in the present studies). The evaporation behavior of deionized water is observed to follow the $D^2$ law (Godsave [7]). Considering the instantaneous droplet diameter $D$, the initial droplet diameter $D_0$ of the droplet, and elapsed time, the law can be expressed as

$$\frac{D^2}{D_0^2} = 1 - k \frac{t}{D_0^2}, \qquad (1)$$

where $k$ represents evaporation rate. The evaporation rate for the water droplet is observed to conform to reports in literature. Figure 2 (a) and (b) illustrates the evaporation behavior for different droplets for variation of field frequency and field strength. Figure 2 (a) shows that increase in frequency from 75 Hz to 225 Hz (at 150 V) leads to improved droplet evaporation rates. Similar to frequency, enhancement in electric field strength also leads to augmented evaporation rates, when the field strength is varied from 0 to 225 V (at 150 Hz), as shown in figure 2 (b).

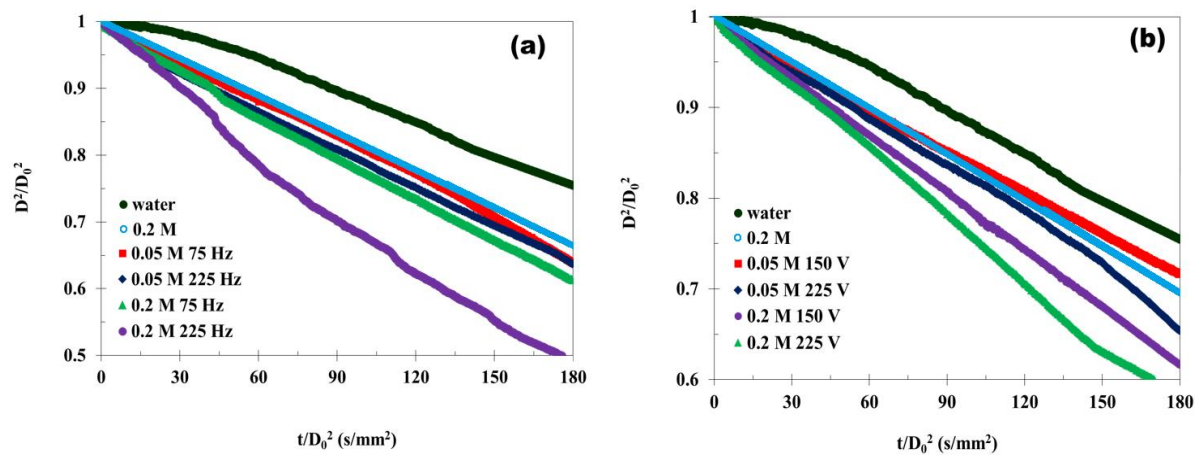

**FIG. 2:** The evaporation behavior of different saline droplets for (a) variation of field frequency at 150 V and (b) variation of field strength at 150 Hz.

The effect of increasing salt concentration has already been probed by the present authors (Jaiswal et al. [8]) and has not been discussed in this article for the sake of brevity. Additionally, the role of salt solubility has also been studied in the mentioned report and accordingly NaI has been selected for the present experiments. It has been reported that the saline droplet experiences internal thermo-solutal advection. The strength of this advection increases with concentration of salt. Due to the internal circulation and the solutal Marangoni advection at the droplet surface, the vapor diffusion layer in contact with the surface of the droplet is also sheared. This shear replenishes the otherwise stagnant diffusion layer with ambient air, leading to improved evaporation [8]. The time lapse arrays in fig. 3 (a) and (b) qualitatively illustrate the role of external field on the . The array illustrates that higher field strengths are more potent towards enhancing the evaporation rates compared to higher field frequencies. At this point, a couple of theories can be proposed as possible mechanisms behind the enhanced evaporation rates in electric field environment. The possible reasons could be modulated surface tension of the conducting fluid due to the Maxwell stresses generated at the droplet interface by the electric field, and the role of electric field induced internal thermohydrodynamics of the droplet. The probable mechanisms shall be discussed in the subsequent sections.

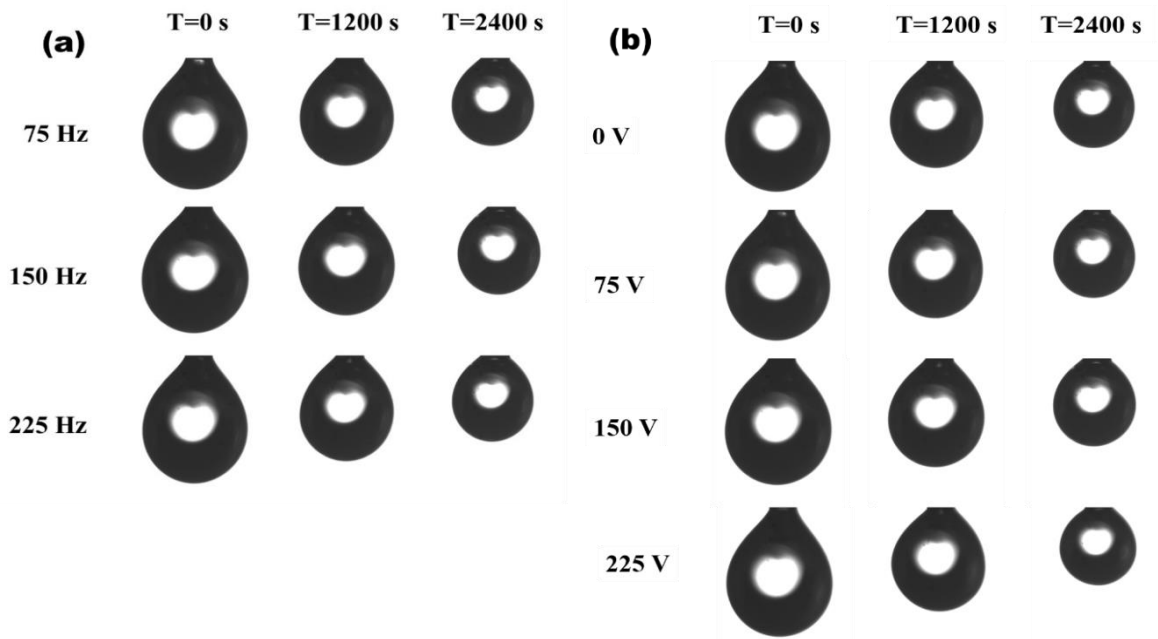

**FIG. 3:** Time-lapse array for the droplets (0.2 M solution) for (a) frequency variation at 150 V and (b) field strength variation at 150 Hz.

## 3. B. Role of surface tension and diffusion dominated evaporation kinetics in electric field environment

The surface tension of a fluid-gas interface is an important physical property which strongly regulates the evaporation rate of the liquid molecules to the ambient gas phase. A lower surface tension dictates that the fluid molecules lodged at the surface require overcoming a weaker energy barrier to escape into the ambient gas phase, and thereby leads to high evaporation or vaporization rates. Further, the surface tension of a fluid, typically conducting fluids such as salt solutions, can be modulated by an electric field due to Maxwell stresses at the surface. The additional stress components at the surface arise due to Lorentz forces on the solvated ions, which remain unbalanced in case of the ions adsorbed at the surface. Consequently, the surface tension of the fluid in field environment must be mapped to understand its role on the improved evaporation. Figure 4 (a) illustrates the nature of variation of the surface tension of 0.2 M NaI solution for increasing field frequency at 150 V and increasing field strength at 150 Hz. The pendent drop method has been employed to determine the surface tension for the fluid. The results have also been further validated by equating the weight of the detached fluid droplet to the surface force at the needle tip during detachment.

It can be observed from fig. 4 that the surface tension of water increases marginally with the addition of salts (0.2 M in this case). The surface tension of the water sample studied has been noted as 72.2 mN/m. Further, with increase of either field frequency (at constant field strength) or field strength (at constant frequency), the surface tension is observe to increase, and then saturate. The increase in surface tension is ~ 5% only. The increment in surface tension due to addition of salts has also been reported in literature [8]. Further, the electric field induces additional Maxwell stresses at the droplets surface due to the solvated ions adsorbed to the interface, which leads to change in surface tension. However, the observations of increased surface tension contradict the notion of improved evaporation rate, and hence is not a plausible mechanism behind the improved evaporation kinetics under electric field stimulus.

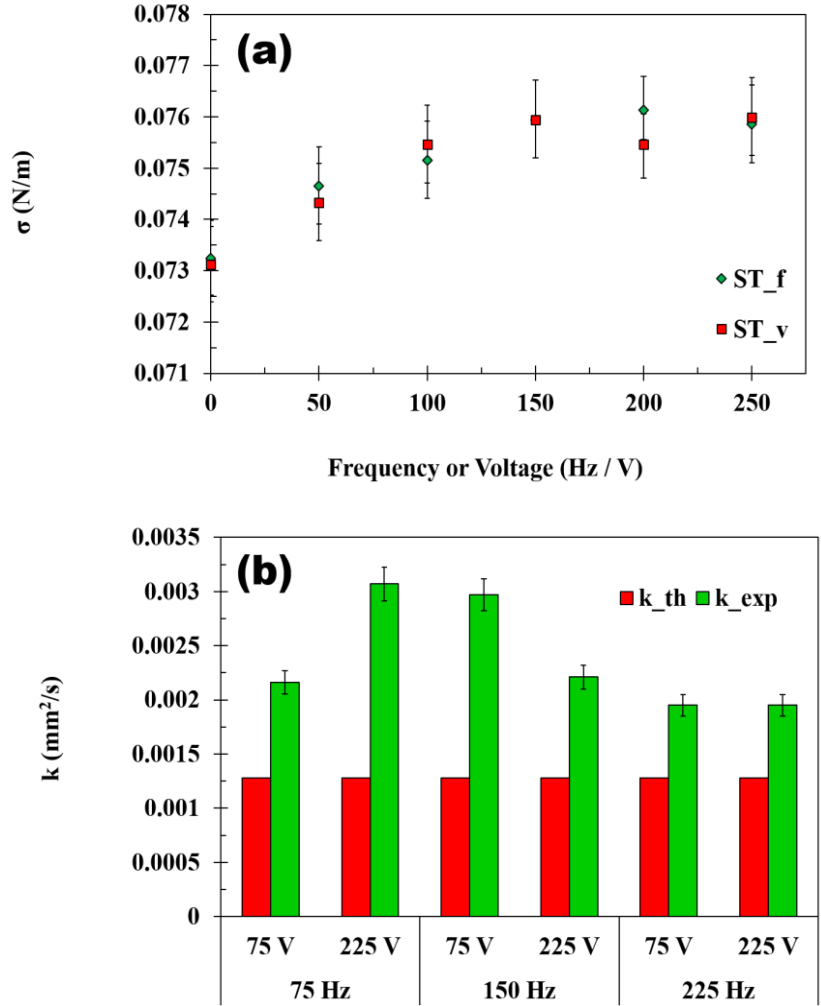

**FIG. 4:** (a) the effect of electric field stimulus on the surface tension of 0.2 M NaI solution. ST_f represents the variation of frequency at 150 V, whereas ST_v represents the effect of field strength variation at 150 Hz, and (b) the comparison of predictions from the diffusion driven evaporation kinetics [38] with the experimentally observed evaporation rates.

For pure fluid droplets in a vapor or gaseous medium, diffusive evaporation to the surrounding environment is the dominant evaporation mechanism for pendant droplets. The classical diffusion driven model by Abramzon and Sirignano [38] can be employed to determine the evaporation rate for pendant droplets to the ambient medium. The model presents the evaporation phenomena as a quasi-steady process, which occurs at the liquid–gas interface. A thin film or shrouding layer of vapor is formed at adjacent to the liquid's outer layer. The vapor diffuses to the ambient gas due to difference in vapor concentration between the vapor shroud and the ambient. The thermal and mass transport occurs through this described thin layer of vapor. Let $Y_s$ and $Y_\infty$ represent the mass concentration of the vapor phase within the shrouding diffusion layer and in the surrounding ambient phase, respectively. Then, Spalding's heat and mass transfer numbers ($B_T$ and $B_M$) are expressible as [38]

$$B_M = \frac{Y_s - Y_\infty}{1 - Y_s} \tag{2}$$

$$B_T = (1 + B_M)^{\frac{C_{pf} Sh}{C_{pg} NuLe}} - 1 \tag{3}$$

In equation 2 and 3, $C_{pf}$, $Sh$, $C_{pg}$, $Nu$ and $Le$ represent the specific heat of the fluid, the associated Sherwood number, the specific heat of the surrounding gas, the associated Nusselt number and the Lewis number, respectively. Based on either eqn. 2 or 3, the mass flux due to evaporation can be deduced as [38]

$$\dot{m} = 2\pi \rho_g D_v R \ln(1 + B_M) Sh = \frac{2\pi \lambda R}{C_{pf}} \ln(1 + B_T) Nu \tag{4}$$

In eqn. 4, $D_v$, $\lambda$, and $R$ represent the diffusion coefficient of the vapor with respect to the ambient phase, the thermal conductivity of the surrounding gas phase and the instantaneous radius of the evaporating pendent droplet, respectively. Based on the theoretical rate of mass evaporation, the rate of decrease of the droplet diameter and subsequently the evaporation rates are assessed. Figure 4 (b) illustrates a comparison between the model predictions and the experimentally observed evaporation rates. It is noteworthy that the model predicts the evaporation rate for water droplet accurate to ±10 % (not illustrated in figure, however, reported in literature by present authors [8] and in other works [37]). The experimental evaporation rate for water is determined as ~ 0.0011 mm$^2$/s, and that of 0.2 M saline solution under zero-field condition is ~0.0017 mm$^2$/s. It can be observed that the theory predicts the same value for evaporation rate for all the cases in fig. 4 (b). This is an expected phenomenon since the diffusion driven model predicts evaporation kinetics based on the gas side conditions, which remains unaltered for all the cases. Therefore, the diffusion driven evaporation kinetics model cannot predict the evaporation rates with field stimulus. Additionally, the surface tension behavior is also not potent to describe the enhanced evaporation rates. Consequently, the mechanism of enhanced evaporation in all probability rests within the droplet and hence the internal transport phenomenon of the droplet during the evaporation needs to be probed. Additionally, figure 4 (b) also shows that at low field frequencies, the field strength plays the dominant role, and the evaporation rate augments with increase in field strength. However, at high frequencies, the dominant role of the field strength no longer holds true, and the evaporation rates remain similar even with increase in field strength. But it is noteworthy that the evaporation at high field frequencies is still higher compared to zero-field case.

**3. C. Nature and influence of internal electrohydrodynamics on the evaporation kinetics**

The internal hydrodynamics of the droplet during evaporation has been observed using particle image velocimetry and the velocity field has been mapped, both qualitatively and quantitatively. PIV studies were carried out for 0.2 M NaI solution droplets, for permutations of field strengths of 75 V and 225 V, and field frequency of 75 Hz and 225 Hz. The PIV setup is validated using internal velocity of evaporating water droplets, and the interior of the droplet is observed to exhibit zero advection, with minor drift motion. This is in agreement with reports on the subject matter [8, 11, 37]. Figures 5 (a)–(d) illustrate the time-averaged velocity contours and vector field for the different electric field configurations mentioned above. For the salt solution droplet at zero-field condition, internal circulation exists (not illustrated, but reported earlier by present authors [8, 11]). The spatio-temporally averaged circulation velocity for this case is ~ 0.185 cm/s (refer fig. 5 (e)). Figure 5 (a) and (b) illustrate the velocity field for frequency variation at 75 V and fig. 5 (c) and (d) illustrate the same for variation of frequency at 225 V. The nature of internal advection within the saline droplet for zero-field case has been reported previously [8, 11], and the plane of circulation corresponds to the plane of the paper (the axis of circulation passing perpendicularly through the paper, and nearly through the center of the droplet). The spatially averaged velocity of advection is as represented in fig. 5 (e).

Based on the PIV studies, it can be inferred that electrohydrodynamic flows within the droplet is active, and the external field tends to enhance the strength of the internal flow. The internal advection within a droplet evaporating in a quiescent ambience can occur due to two typical possibilities, viz. the thermal and solutal gradients within the droplet. It is noteworthy that the solutal gradient is possible on in case of complex fluid (salt solutions, colloids, surfactant solutions, etc.), binary fluid droplets, etc. the evaporation process leads to cooling of the droplet's interior compared to its surface, which generates a thermal gradient across the droplet. Likewise, the pendent shape promotes unequal evaporation along the droplet; leading to thermal gradients generated along the droplet surface as well. This gives rise to minor (for water droplets) thermal advection within the droplet as well as thermal Marangoni circulation at the droplet interface (due to the temperature dependence of surface tension) [8, 37].

Along similar lines, saline droplets exhibit solutal advection and solutal Marangoni circulation (due to the solute concentration dependence of surface tension). Solvated ions due to addition of salts are preferentially adsorbed-desorbed at the droplet surface, which is evident from the modulated surface tension of the fluid (refer fig. 4 (a)). The evaporation from the surface leads to a concentration gradient between the droplet's interior and interface. Additionally, the non-uniform evaporation from the pendent shape leads to solutal Marangoni circulation at the droplet interface [8, 11]. The advection currents generated within the droplet and at the surface of the droplet develops shear at the liquid-gas interface. This shears the vapor diffusion layer surrounding the droplet, and replenishes the otherwise stagnant layer with ambient air. This prevents the diffusion layer from saturating with vapor and improves the evaporation rate [8, 11, 37].

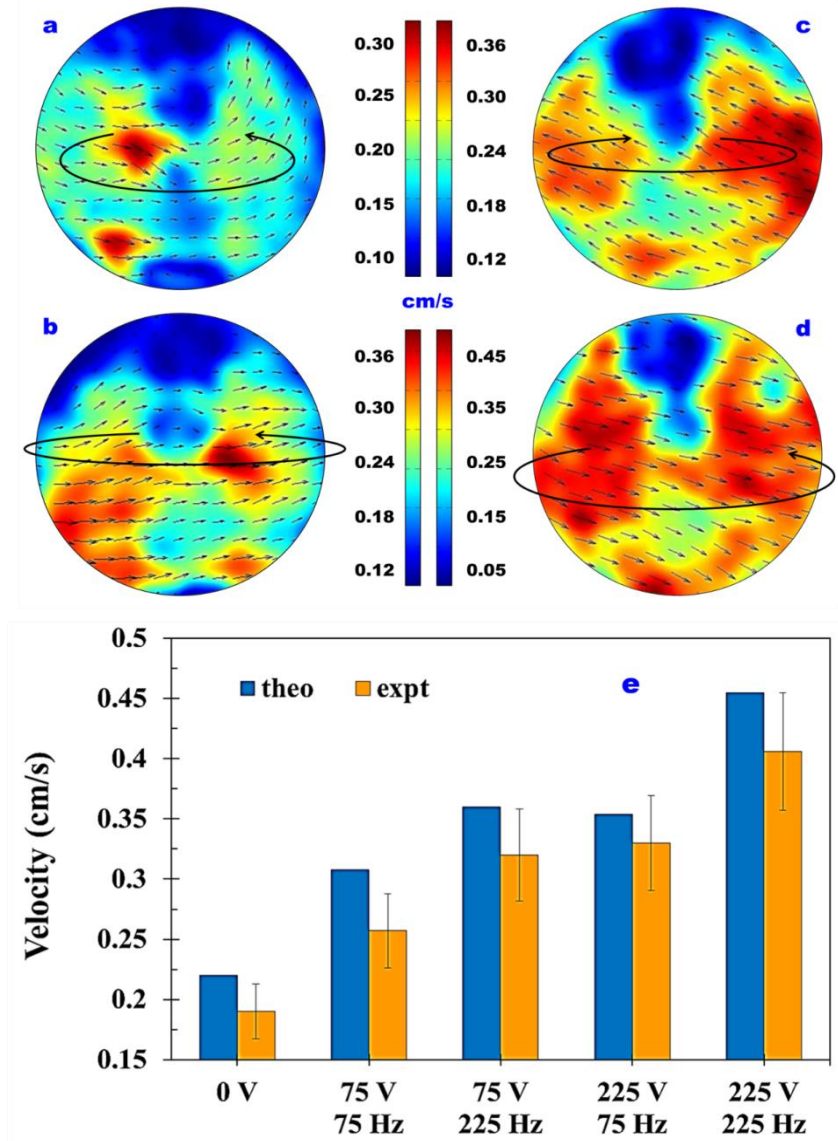

**FIG. 5:** Time-averaged mean velocity contours within 0.2 M NaI solution droplets with different field constraints, as (a) 75 V and 75 Hz (b) 75 V and 225 Hz (c) 225 V and 75 Hz and (d) 225 V and 225 Hz. (e) Illustration of the comparison of theoretical velocities (of the spatio-temporally averaged velocities) with respect to the velocities predicted from the present mathematical model.

### 3. D. Role of electro-thermal advection

The PIV study has established that electric field mediated internal advection exists within the evaporating saline droplet. Further, as discussed, such motion may be caused by either thermal gradients or solutal gradients. Hence, it is important to determine the governing mechanism behind such internal advection. It is assumed that the thermo-physical properties remain invariant over the small range of temperature and humidity gradients that are generated due to the evaporation process. Additionally, the same holds good with the application of the electric field. However, the surface tension varies due to the thermal and solutal gradients established within the droplet during evaporation. Considering the electro-thermal effect as the dominant mechanism behind the internal advection, the energy balance can be expressed mathematically as

$$\dot{m} h_{fg} = k_{th} A \frac{\Delta T_m}{R} + \rho C_p U_c A \Delta T_m + \rho C_p V_f A \Delta T_m \tag{5}$$

In eqn. 5, the left hand side represents the energy associated with the evaporative mass flux from the droplet surface. This is balanced by the thermal energy transport within the droplet, and the terms on the right hand side represent the conduction transport, convective transport due to internal flow, and the convective transport due to the electro-thermal flows, respectively. In eqn. 5 $\dot{m}$ and $h_{fg}$ represent the evaporative mass flux and the enthalpy of vaporization. The variables $k_{th}$, $A$, $\rho$, $\Delta T_m$, $C_p$, $U_c$ and $V_f$ represent the thermal conductivity of the fluid, the area of cross section of the droplet, the density of the fluid, the temperature difference between the interface and the bulk of the droplet, the specific heat capacity of the fluid, the mean internal circulation velocity due to thermal advection, and the mean internal circulation velocity due to electric field driven advection, respectively.

The advection velocity due to thermal gradient is deduced as $U_c = \sigma_T \Delta T_m / \mu$, where $\sigma_T$ and $\mu$ represents differential surface tension with respect to temperature change, and viscosity of the fluid [8, 37]. The solvated ions in the polar liquid are preferentially adsorbed-desorbed to the liquid-gas interface. The thermal advection leads to a system of moving solvated charges within the liquid. In the presence of an external electric field, the solvated ions give rise to a body force, which leads to an additional electro-thermal advection component, leading to enhanced circulation velocity. The electro-thermal advection velocity $V_f$ is obtained by balancing the electrohydrodynamic forces in the system. The generic electromagnetic force on a charged system due to imposed fields can be expressed as

$$\vec{F} = q\vec{E} + (\sigma_e \cdot \vec{E} \times \vec{B}) + \sigma_e (\vec{v} \times \vec{B}) \times \vec{B} \tag{6}$$

In eqn. 6, $\vec{F}, q, \vec{E}, \vec{B}, \sigma_e$ and $\vec{v}$ represent the electromagnetic force, the net charge of the ionic population, the electric field strength, the magnetic field strength, the electrical conductivity of the fluid, and the velocity of the charged particles, respectively. In the present case, no magnetic field is involved, and eqn. 6 reduces to

$$\vec{F} = q\vec{E} \qquad (7)$$

The force per unit volume (*f*) is hence expressed as

$$f = \rho_e \vec{E} \qquad (8)$$

where $\rho_e$ is the charge per unit volume, which is further expressed as $\rho_e = zeN$. Here z, *e* and *N* represent the valence of the solvated ions, the magnitude of elementary charge and the number density of solvated ions in the system. The electric body force leads to inertia within the electrohydrodynamic system as

$$\rho_e E = \rho a \qquad (9)$$

In eqn. 9, *a* represents the acceleration due to inertial circulation within the droplet. The acceleration component can be scaled as $\sim V_f/t$ and further, the time elapsed can be scaled as $t \sim R/V_f$, where *R* is the instantaneous radius of the droplet. Substitution of the scaled acceleration term *a* in eqn. 9 yields

$$V_f = \sqrt{\frac{\rho_e E R}{\rho}} \qquad (10)$$

Further, $V_f$ can be simplified in terms of the electric field potential ($V=ER$) and the volumetric charge density as

$$V_f = \sqrt{\frac{eNzV}{\rho}} = \frac{v}{R}\sqrt{\frac{eNzVR^2}{\rho v^2}} \qquad (11)$$

$$V_f = \frac{v}{R}\sqrt{E_{HD}} \qquad (12)$$

In eqn. 12, $\nu$ represents the kinematic viscosity and $E_{HD}$ represents the associated Electrohydrodynamic number. Upon substitution of all the obtained expressions in eqn 5, the latter is expressible as

$$\rho \dot{R} A h_{fg} = k_{th} A \frac{\Delta T_m}{R} + \rho C_p A \Delta T_m \frac{\sigma_T \Delta T_m}{\mu} + \rho C_{pf} A \Delta T_m \frac{\nu}{R} \sqrt{E_{HD}} \quad (13)$$

$$\rho \dot{R} R h_{fg} = k_{th} \Delta T_m [1 + \rho C_p \frac{R \sigma_T \Delta T_m}{k_{th} \mu} + \rho C_{pf} \frac{\nu}{k_{th}} \sqrt{E_{HD}}] \quad (14)$$

The terms in eqn. 14 can be further simplified and rearranged to yield

$$\rho \dot{R} R h_{fg} = k_{th} \Delta T_m [1 + Ma_T + \Pr \sqrt{E_{HD}}] \quad (15)$$

where $\Pr$ and $Ma_T$ represent the associated Prandtl number and thermal Marangoni number, respectively. For stable Marangoni circulation, the Ma must be over and above a critical value (~80), [37] and hence eqn. 15 can be simplified as

$$\rho \dot{R} R h_{fg} = k_{th} \Delta T_m [Ma_T + \Pr \sqrt{E_{HD}}] \quad (16)$$

The second term within brackets in the RHS of eqn. 16 represents the effect of the electric field on the thermal diffusion and is termed as the electro-Prandtl number. The effective electro-thermal Marangoni number governing the process can be expressed as $Ma_{T,e} = Ma_T + \Pr\sqrt{E_{HD}}$, which represents the net modulation of the thermal Marangoni advection due to presence of the electric field.

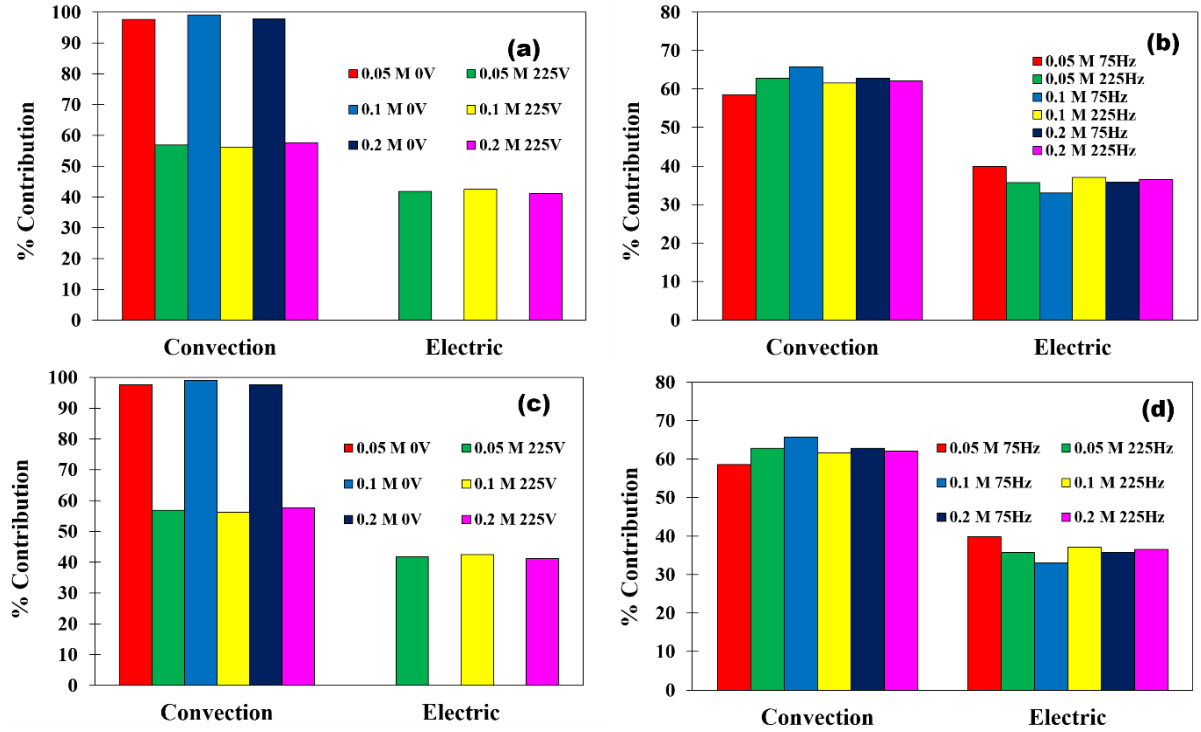

**FIGURE 6:** The contribution of advection and electrohydrodynamic advection components in thermal transport with (a) frequency variation at 150 V and (b) field strength variation at 150 Hz, and in species transport with (c) field strength variation at 150 Hz and (d) frequency variation at 150 V.

Figure 6 a and b illustrate the effective contribution of thermal diffusion, thermal advection and electro-thermal advection (the terms on RHS of eqn. 13), with variation of frequency and electric field strength. The figure does not include the contribution due to thermal diffusion, as it is insignificant (~1-2 %) when compared to the other mechanisms. For example, for 0 V case of a saline droplet, the contribution due to the thermal advection component is determined as ~ 97 %, and only ~ 3 % is contributed by thermal diffusion. Hence, under zero-field condition, the thermal advection is dominant over the thermal diffusion mechanism within the droplet. In the presence of electric field, the electro-thermal advection term is introduced to the energy conservation equation. It is observed that upon deducing the magnitudes, the contribution of the electro-thermal advection term is typically 40-45 %, with a reduction in the thermal advection component at zero-filed. Thereby, the scaling analysis reveals that a large proportion of the internal advection within the droplets due to electric field is caused by the electrohydrodynamics and electro-thermal effects of the ion solvated polar fluid.

With the role of thermal and electro-thermal advection being discussed, the genesis of the same needs to be probed. Thermal advection may be a typical Rayleigh advection, which is brought forth by thermal gradient induced adverse density gradients within the system. It could also be due to the thermal Marangoni advection, which is prominent in such droplets and liquid films. Similar to the advection velocity due to thermal Marangoni, the advection velocity due to buoyancy effects is expressible as [37]

$$u = g\beta \Delta T_R R^2 / \nu \tag{17}$$

In eqn. 17 $g$ and $\beta$ represent the acceleration due to gravity, and the volumetric thermal expansion coefficient of the fluid, respectively. The temperature difference $\Delta T_R$ within the droplet, due to which buoyancy advection effect might arise is expressible as [37, 8]

$$\Delta T_R = \sqrt{\frac{\nu \dot{R} h_{fg}}{g \beta R^2 C_p}} \tag{18}$$

Where, $\dot{R}$ is the rate of change of droplet radius. The associated Rayleigh number $Ra$ is expressed as

$$Ra = \frac{R^2}{\alpha} \sqrt{\frac{\dot{R} h_{fg} g \beta}{C_p \nu}} \tag{19}$$

To determine the dominant cause for the internal thermal advection, the analysis of Rayleigh-Marangoni advection and approach by Nield [39] and Davis [40] has been adopted. The model evaluates the relative dominance of the Rayleigh and Marangoni convection in a conjugate system, and their strength is deduced with respect to their respective critical values (determined from stability analysis). The mathematical expression for the same is as per eqn. 20.

$$\frac{Ra}{Ra_c} + \frac{Ma_T}{Ma_{T,c}} = 1 \tag{20}$$

Where, $Ra_c$ and $Ma_{T,c}$ represent the critical values of the associated Rayleigh number and thermal Marangoni number When plotted on a Ma-Ra plane, the relative dominance of the two modes can be determined using the criteria by Nield [39] and Davis [40]. The critical Ra is set to 1708 (as per Chandrasekhar's stability criteria), while the critical Ma is set to 80 (by

Nield's analysis) [41]. Further, Davis put forward a regime of marginally-stable internal advection, governed by a critical Ma of ~ 53-54. Figure 7 (a) illustrates the map of thermal Marangoni number ($Ma_T$) and $Ra$ with variation of field frequency at 150V and 7 (b) illustrates the same for variation in field strength at 150 Hz, respectively.

For pure water, the point lies in the zone of unstable advection, which signifies that the droplet exhibits no internal hydrodynamics, and this is in agreement with experimental observations. The graph contains three regions, the region below the Davis line, which represents region of no internal advection, the region between the Davis and Nield lines, which represents the zone of partially stable and intermittent internal circulation, and the region above the Nield line, which represents the zone of stable internal circulation. As observable in fig. 7 (a) and (b), the presence of salt in the system enhances the values of both the $Ma_T$ and Ra; however, the regime still lies in the region of unstable internal circulation. The increase of either field frequency or the field strength shifts the regimes towards the right and the upper side, but the advection still remains within the region of unstable circulation. However, this is in contrast to the experimental observations, where the electric field induced internal advection is stable and persists over time, signifying stable or partially stable circulation. Thereby, the $Ma_T$ vs Ra approach is not physically consistent and a different approach is essential describe the mechanism.

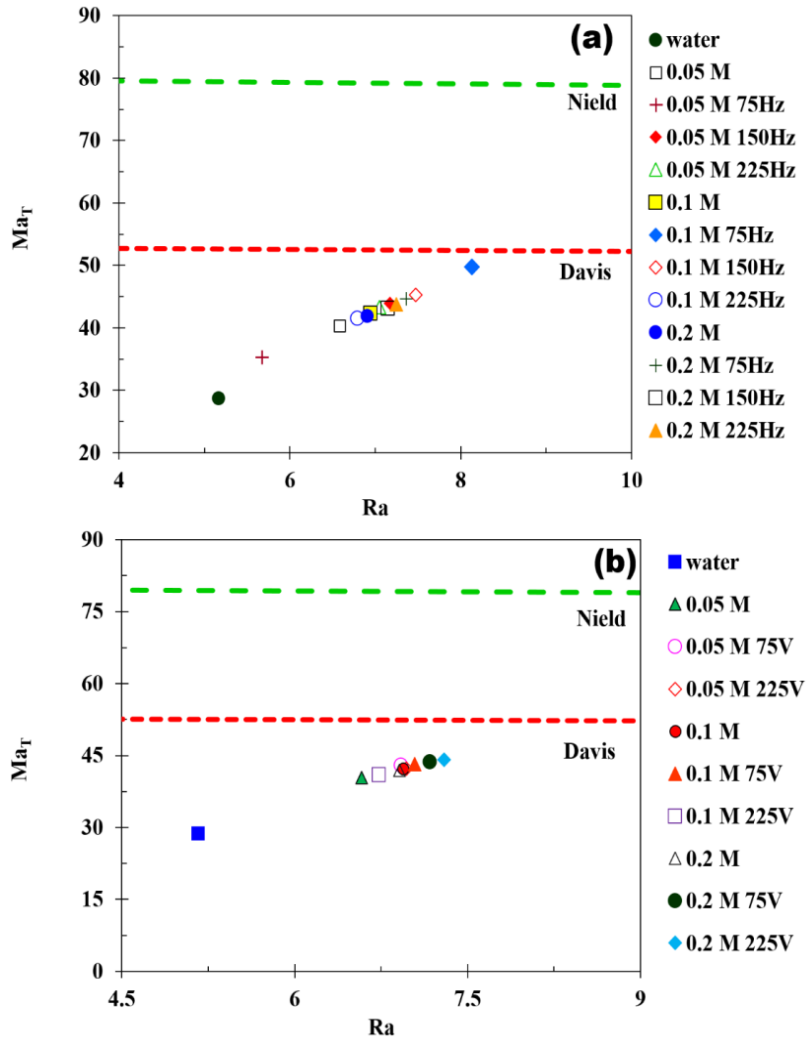

**FIGURE 7:** Stability plot for thermal Ma against Ra to determine the dominant mode of internal thermal circulation, with variation of (a) field frequency (b) field strength. The two lines represent the stability regions by Nield [39] and Davis [40]. The entries in the legend without any field parameters represent zero-field cases.

Consequently, the effective electro-thermal Marangoni number, $Ma_{T,e}$ has been plotted against $Ra$ to determine the stability of the internal advection. The variation the $Ma_{T,e}$ against the Ra, for different salt concentration and field constraints have been illustrated in fig. 8 (a) and (b). At zero-file, the $Ma_{T,e}$ is equal to the $Ma_T$, and hence lies in the region of unstable advection. However, in the present case, with increase in field parameters, all the points shift to the region of partially stable advection. This is in agreement with experimental observations, where significantly improved internal circulation behavior is noted with the application of electric field. This further cements the hypothesis that electro-thermal

Marangoni advection is a key component behind the augmented internal advection, which leads to improved evaporation rate. In addition to the stability regimes by Nield and Davis, the iso-$E_{HD}$ lines have also been illustrated in the stability map to represent the behavior of the electrohydrodynamics within the droplets with different field constraints. It is observable that with increasing $E_{HD}$ at a constant Ra, the iso-$E_{HD}$ lines shift upwards, signifying increasing contribution to internal circulation by the electro-thermal Marangoni effect. Further, it is also noteworthy that the separation between the iso-$E_{HD}$ lines reduces with its increasing magnitude. This essentially signifies that with increasing field strength or frequency, the strength of the internal advection gradually saturates and reaches a plateau, where stable circulation is consistent.

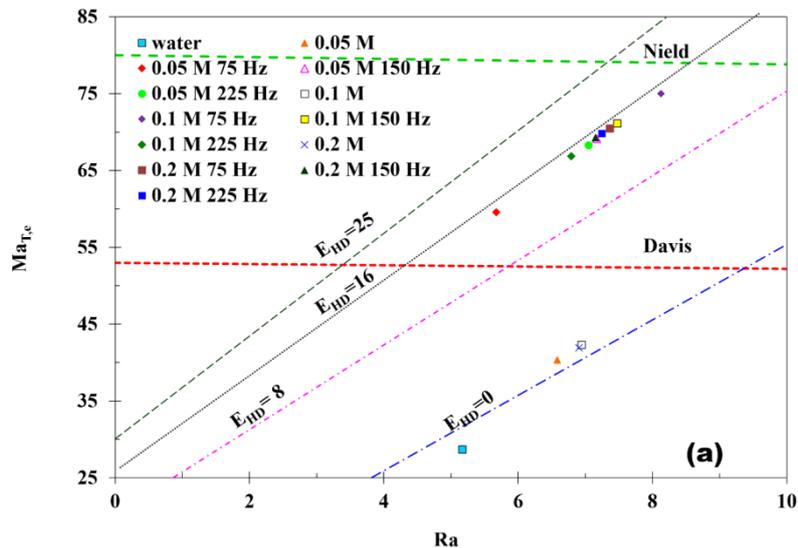

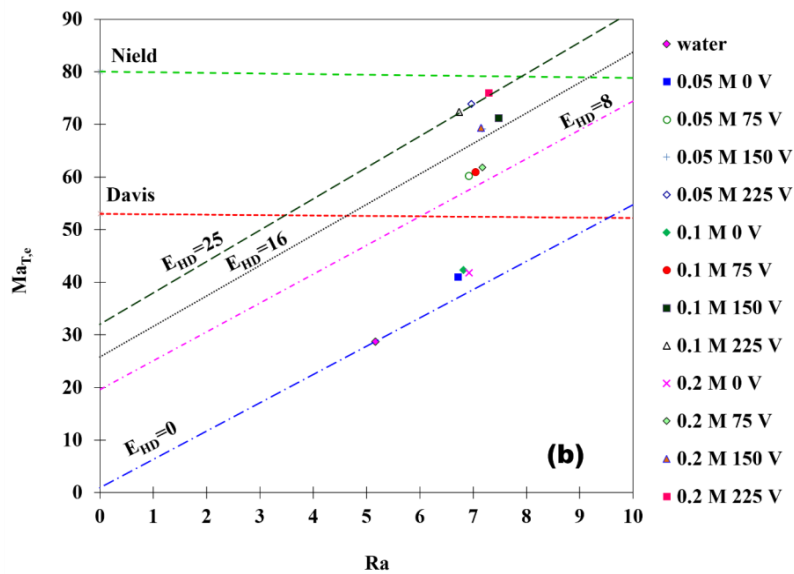

**FIGURE 8**: Stability plot for electro-thermal Marangoni number ($Ma_{T,e}$) against Ra with variation of (a) the field strength and (b) field frequency. The different lines cutting across the plot represent the iso-$E_{HD}$ lines, which represent the various saline droplets and electric field constraints non-dimensionally, and act as stability lines for the electro-thermal Ma system. The entries in the legend without any field parameters represent zero-field cases.

### 3. E. Role of electro-solutal advection

The thermal Marangoni advection within evaporating droplets is largely common for fluids with temperature dependent surface tension, with water, alcohols and colloids being typical examples [42]. However, surface tension of polar fluids is also strongly dependent on the concentration of solvated solutes or of other miscible fluid molecules. Hence, the inclusion of salts in water also introduces the solutal Marangoni effect and its response to change in salt concentration due to evaporation further modulates this advection behavior. Similar to the thermal Marangoni effect, the shape of the pendent induces non-uniform evaporation from the droplet, leading to concentration gradient along the droplet surface, which leads to the solutal Marangoni advection on the droplet surface. Additionally, solvated ions are preferentially adsorbed to the droplet interface compared to the bulk of the droplet (as evident from the change in surface tension of the saline fluid). The dynamic change in solute concentration due to evaporation, within the droplet bulk as compared to the droplet's interface, gives rise to the solutal internal advection. Works have shown [8, 42] that in salt solution droplets, the augmentation in the evaporation rate is dominated by the solutal advection compared to the thermal advection. Consequently, the role of the electro-solutal advection, along similar lines of the electro-thermal advection, also needs to be probed to obtain a clear picture of the physics involved.

Solvated ions preferentially adsorb-desorb to the liquid-gas interface with respect to the bulk of the liquid, which is manifested as the change in surface tension upon addition of salts. Consequently, the interfacial concentration and the bulk concentration of solvated ions in a droplet are not equal, which gives rise to the solutal advection. Further, evaporative flux occurs from the surface of the droplet, which alters the surface concentration of solvated ions at the surface in a stronger manner than the ion concentration at the bulk. This aggravates the concentration difference between the bulk of the droplet and the surface, which further improves the solutal advection inside the droplet. The dynamic bulk concentration can be determined by conservation of the total species present in the system. The product of V (instantaneous volume) and C (instantaneous bulk concentration) is conserved throughout the evaporation of the droplet ($VC = constant$), which is employed to determine the instantaneous bulk concentration from the instantaneous volume of the droplet. The dynamic interfacial concentration is determined from the temporally evolving surface tension, since the surface tension of the saline droplet is dependent on concentration (refer fig 4) [8]. The temporal variation of the surface tension is quantified by fitting instantaneous droplet shape profiles to the Young-Laplace equation.

The time dependent evolution of surface tension is expressed in the form of an equation from the obtained experimental data. Likewise, the dependence of surface tension on the salt concentration and field constraints is expressed in terms of an equation. The two equations are employed to obtain the expression for dependence of interfacial concentration with time. It is noteworthy that these relationships were deduced for the initial 15-20 minutes to ensure that the governing Worthington number does not drop below ~0.6 [41] which ensures that the drop shape analysis is proper. With evolving time, the Worthington number ($Wo = V/V_0$, where $V_0$ is the initial volume) reduces and renders the shape fitting exercise prone to errors. The dynamic bulk and interfacial concentrations for different saline droplets and electric field constraints have been illustrated in fig. 9 (a) and the difference between the (b). It is observable that under any constraint, a difference exists between the bulk and the interfacial ionic concentrations, which establishes that solutal advection is also present within the droplet. Additionally, the strength of the solutal advection increases with time due to the increasing difference in concentration with time.

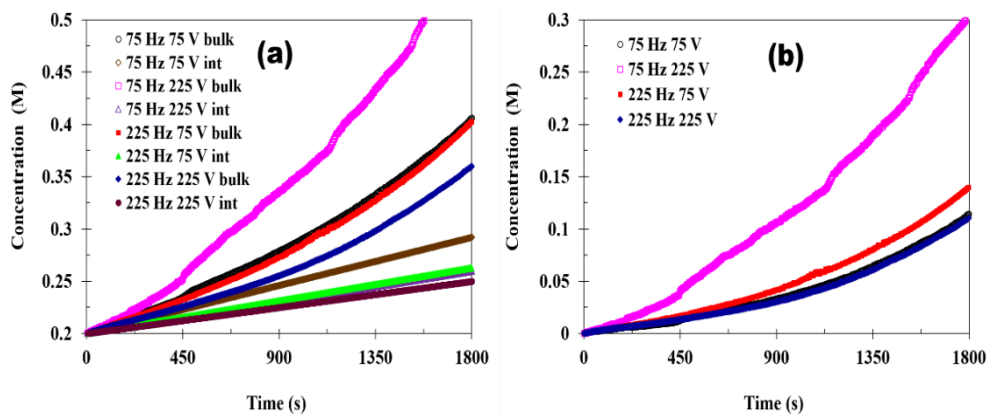

**FIGURE 9:** (a) The temporal variation of the dynamic and bulk concentration within the droplet (0.2 M NaI) for different field strengths (75 and 225 V) and frequencies (75 and 225 Hz). The term '*int*' represents the dynamic interfacial concentration, and '*bulk*' represents the dynamic bulk concentration (b) The difference between the bulk and interfacial concentration for different field constraints.

Scaling the species transport mechanisms, the mass lost due to evaporation can be deemed responsible for the species diffusion transport, solutal advection and electro-solutal advection (only in presence of electric field) within the droplet. Mathematically, the scaled equation is represented as

$$\dot{m} = DA \frac{\Delta C_m}{R} + U_{c,m} \Delta C_m A + V_{f,c} \Delta C_m A \tag{21}$$

Where, $\dot{m}$, $D$, $\Delta C_m$, $U_{c,m}$ and $V_{f,c}$ represent the rate of evaporative mass flux from the droplet, the diffusion coefficient of the salt with respect to water, the solute concentration gradient present within the droplet, the velocity of internal circulation caused by the solutal advection and the velocity of electro-solutal advection in the presence of electric field, respectively. The velocity of solutal advection is expressed as $U_{c,m} = \sigma_c \Delta C_m / \mu$, where $\sigma_c$ represents the rate of change of surface tension with respect to change of solute concentration [8, 42]. Similar to the electro-thermal analysis, the $V_{f,c}$ is deduced as

$$V_{f,c} = \frac{D}{R} \sqrt{E_{HD}} \tag{22}$$

Substituting the expressions for the advection velocities in eqn. 21, eqns. 23 and 24 are obtained

$$\rho \dot{R} R = D \Delta C_m + \frac{\sigma_c (\Delta C_m)^2 R}{\mu} + \nu(\sqrt{E_{hd}}) \Delta C_m \tag{23}$$

$$\frac{\rho \dot{R} R}{D \Delta C_m} = (1 + Ma_s + Sc\sqrt{E_{hd}}) \tag{24}$$

In Eq. 24, $Ma_s$ represents the solutal Marangoni number and $Sc$ represents the Schmidt number. For Marangoni advection to manifest, Ma >>1 [37], and hence the expression can be simplified as

$$\frac{\rho \dot{R} R}{D \Delta C_m} = Ma_s + \sqrt{Sc_{sp}} \sqrt{E_{hd} Sc_i} \qquad (25)$$

The right-hand side term $\left( Ma_s + \sqrt{Sc_{sp}} \sqrt{E_{hd} Sc_i} \right)$ essentially behaves as the effective electro-solutal Marangoni number $(Ma_{s,e})$ for the system. The eqn. 24 has been remodeled to express the effective Sc as composed of the $Sc_{sp}$, which represents the species transport Schmidt number and the $Sc_i$, which represents the ionic Schmidt number. The $Sc_{sp}$ governs the strength of the momentum diffusion created due to internal circulation to the mass diffusion of the solvated ions. The $Sc_i$ on the contrary is the ratio of the ionic mobility $(m_{ion})$ to the diffusivity of the solvated ions $(D_{ion})$, and is important during the electro-solutal advection. Figure 6 (c) and (d) illustrate the effects of variation of field strength and frequency on the different modes of species transport described by eqn. 21. Upon comparison with fig. 6 (a) and (b), it can be observed that the degree of contribution of the electro-thermal advection (with respect to the zero-field thermal advection) is nearly similar to the degree of contribution of the electro-solutal contribution (with respect to the zero-field solutal advection).

Hence, in the present case it can be theorized that both the electro-thermal and electro-solutal advection contributes largely to the improved internal circulation, which eventually leads to increase in the evaporation rates. This is in stark comparison to the equivalent magnetohydrodynamics within droplets, where the magneto-solutal effect dominates over the magneto-solutal effect [11]. The stability of the electro-solutal Ma advection can be determined from an analogous stability map of the Ma$_{T,e}$ and the Ma$_{S,e}$, as illustrated in fig. 10. Fig. 10 (a) illustrates the phase map for the $Ma_T$ against the $Ma_s$, in the absence of electric field. All the points in fig. 10 (a) lie to the right of the iso-Lewis number line, $Le = 0$ (not illustrated in figure) (Joo [43]). This represents that the solutal advection within the system is stable and the circulation within the droplet due to solutal gradient is a stable circulation. However, the relative roles played by the electro-thermal and electro-solutal Marangoni effects are not clear from this illustration. Fig. 10 (b) illustrates a phase plot of the effective thermal Marangoni number against the effective solutal Marangoni number for field strength and frequency variation. With the electro-diffusive term incorporated in the effective electro-thermal and electro-solutal Ma, the points shift further away from the origin, towards the right and upwards, which indicates increment of both the thermal and solutal Marangoni advection towards even more stable circulation. This nature of the shift is also evident from the nature of the iso-$E_{HD}$ lines, which have been included to map the stability of the field mediated internal circulation.

However, the relative increment in the $Ma_{T,e}$ with respect to the $Ma_T$ and the relative increment of the $Ma_{S,e}$ with respect to the $Ma_S$ are of similar magnitude. In case of no-field, the solutal Ma is the dominant cause of advection compared to the thermal Ma [8]. However, as revealed by the present analysis, the augmentations in the effective electro-solutal Ma and the effective electro-thermal Ma are very similar in magnitude. Hence, in case of electrohydrodynamics mediated improvement in evaporation of pendent droplets, both the electro-thermal and electro-solutal advection play nearly equal important roles. Since both the electro-thermal and electro-solutal advection are noted to be of similarly dominant, the internal advection velocity can be deduced as a mean of both approaches (eqns. 12 and 22). The mean advection velocities determined from the theoretical approach have been compared against representative experimental observation in fig 5 (e). Good agreements have been noted, which further strengthens the proposed mechanisms.

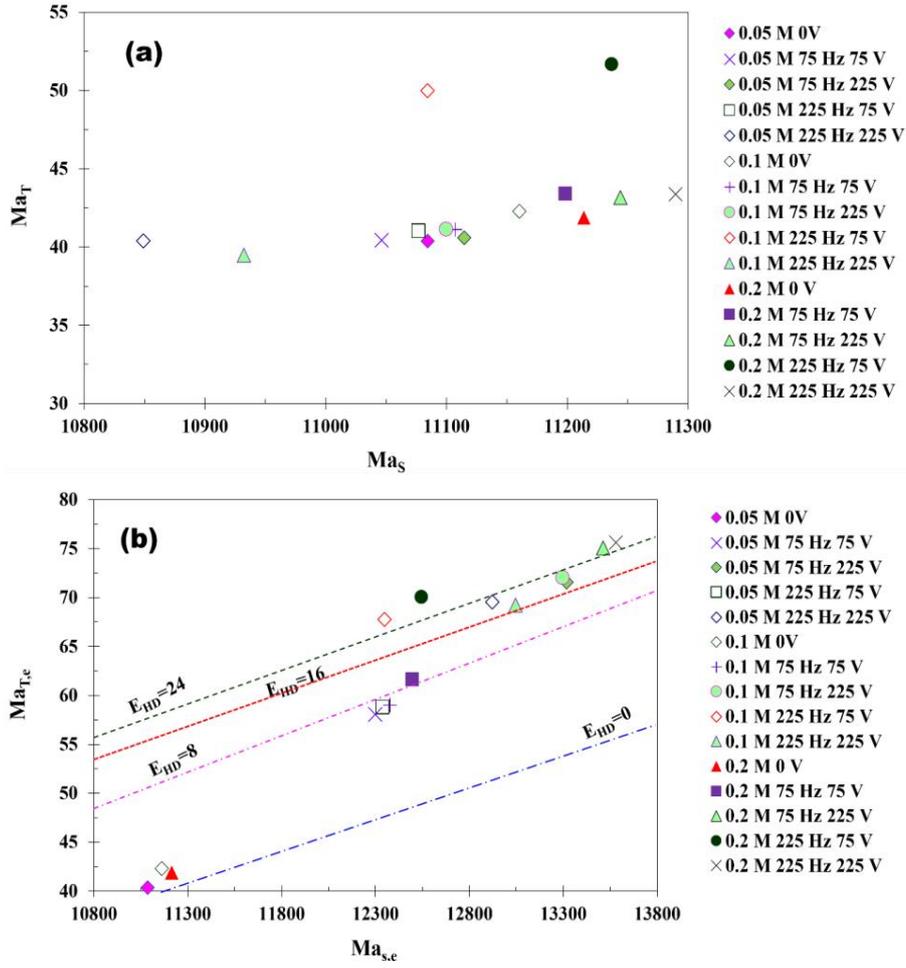

**FIGURE 10:** The phase map of the (a) thermal Ma against the solutal Ma in the absence of electric field (b) effective electro-thermal Ma against the effective electro-solutal Ma with electric field under variant field constraints. The thermo-solutal phase map is adapted from literature (Joo [43]). The entries in the legend without any field parameters represent zero-field cases.

## 4. Conclusions

The present article studies and discusses the effects of electric field on the evaporation characteristics of pendent droplets of conducting fluids. For the first time in literature, the improvement of evaporation of a droplet of a conducting fluid by the presence of an electric field, has been experimentally studied and theoretically verified and explain. The evaporation kinetics of saline pendent droplets was experimentally studied and the evaporation rate was observed to increase with the application of an external electric field. The degree of increase of evaporation rate was found to hold a direct relationship with the increase of field strength and field frequency. All the droplets under different field constraints are observed to conform to the classical $D^2$ law, which signifies that the vapor side is unaffected by the electric field. Probing the plausible cause for the improved evaporation rate reveals that the role of surface tension cannot be employed to explain the mechanism, as contradictory behavior is observed. Further, the classical diffusion driven evaporation model is found to fail in predicting the evaporation rate constant under electric field constraints.

Consequently, it is evident that the gas side is not directly responsible towards the augmented evaporation, and the liquid side is probed using PIV. The study reveals that the thermo-solutal advection within the droplet is largely augmented by the presence of the electric field. This enhanced internal circulation velocity thus strongly shears the droplet interface, which further shears the vapor diffusion layer shrouding the droplet. This replenishes the vapor diffusion layer with ambient air, thereby improving the species transport from the droplet. Having identified the cause behind the improved evaporation, the mechanisms behind the same have been investigated. Theoretical scaling models based on electro-thermal and electro-solutal Marangoni advection has been put forward. The models reveal that the electro-thermal and electro-solutal Marangoni effects and the field mediated internal electro-thermo-solutal advection lead to the improved circulation. It is also shown that the electro-thermal and electro-solutal advection lies in the regions of stable advection on the respective stability maps. It is possible to determine the internal advection velocities from the proposed scaling and good agreements with experiments have been noted. Further, it is noted that both the modes of transport are nearly equally dominant towards the improved evaporation behavior. The present findings may find strong importance in microscale electrohydrodynamic systems.

## Acknowledgments

PD acknowledges the funding by IIT Ropar for the present research (vide grants IITRPR/Research/193 and IITRPR/Interdisp/CDT).

## Conflict of interests